\documentclass[journal=jacsat,manuscript=communication]{achemso}
\usepackage{txfonts}

\usepackage[version=4]{mhchem}

\author{Ling Liu}
\author{Yuyang Liu}
\author{Chungen Liu}
\email {cgliu@nju.edu.cn}
\affiliation{Institute of Theoretical and Computational Chemistry,
	Key Laboratory of Mesoscopic Chemistry of the Ministry of Education (MOE),
	School of Chemistry and Chemical Engineering,
	Nanjing University,
	Nanjing, 210023,
	China }

\title{Enhancing the understanding of hydrogen evolution and oxidation reaction on Pt(111) through ab initio simulations on electrode/electrolyte kinetics}


\begin{document}
	
\begin{abstract}
The hydrogen oxidation reaction (HOR) and hydrogen evolution reaction (HER) play an important role in hydrogen-based energy conversion. Recently, the frustrating performance in alkaline media raised debates on the relevant mechanism, especially on the role of surface hydroxyl (\ce{OH}$^*$). We present a full pH range electrode/electrolyte kinetics simulation for HER/HOR on Pt(111), with the potential-related rate constants been calculated with density functional theory methods. The polarization curves agree well with the experimental observations. The stability of \ce{OH}$^*$ is found to be unlikely an effective activity descriptor since it is irrelevant to the onset potential of HOR/HER. Degree of rate control analyses reveal that the alkaline current is controlled jointly by Tafel and Volmer steps, while the acidic current solely by Tafel step, which explains the observed pH-dependent kinetics. Therefore, it is also possible to reduce the overpotential of alkaline HER/HOR by accelerating the Tafel step besides tuning the hydrogen binding energy.
\end{abstract}

A promising direction of utilizing the hydrogen energy is through combining the water electrolysis as well as the hydrogen fuel cell. The hydrogen evolution reaction (HER) and its reverse, the hydrogen oxidation reaction (HOR) are two relevant electrochemical reactions in aqueous environment. Although platinum and its group metals rank among the most active electrocatalysts for HER/HOR, the sluggish kinetics in alkaline conditions has hindered their application in alkaline electrolysers and anion exchange membrane fuel cells.\cite{Durst2014,Sheng2010} The dramatic pH effect of HER/HOR has been rationalized as the difficulty of water dissociation,\cite{Strmcnik2013,Zheng2018} or the variation of transport rate of \ce{OH-}/\ce{H+},\cite{Koper2013,Rossmeisl2016} or the shifting of apparent hydrogen binding energy (HBE)\cite{Cheng2018,Sheng2015}. The activity descriptor for alkaline HOR/HER is also controversial. As HBE is known as the sole descriptor for HER/HOR processes in acidic environment,\cite{Trasatti1972,Norskov2005,Skulason2010} the adsorbed \ce{OH}$^*$ was either suggested as the second descriptor for alkaline mechanism,\cite{Strmcnik2013,Koper2013,Strmcnik2016} or an inhibitor blocking the active sites.\cite{Schmidt2002,Markovic2002,Intikhab2017} Contrarily, much less theoretical works have been directed towards the understanding of alkaline mechanism, which should reveal the inherent reasons underlying the alkaline activity at the atomic level. 

Recently, the development of theoretical methods in calculating the potential-related activation barriers in electrochemical interface has made great progress,\cite{Chan2015,Goodpaster2016,Cheng2016,Xiao2017a} and the combining with microkinetics simulation could provide a reliable way to verify the proposed mechanisms and go to in-depth exploration of the reaction characteristic.\cite{Hansen2014,Singh2017} To clarify the on-going debates, we present here the first multiscale simulation of polarization curves for the full pH range of HER/HOR on Pt(111) by solving the integrated equations of the diffusion-layer particle transport and the metal/electrolyte interfacial microkinetics. The potential-related kinetics parameters of the interfacial elementary steps are computed with our previously proposed scheme.\cite{Liu2018} The energies are evaluated with density functional theory (DFT), while the potential and solvation effects are approximated by an implicit solvation model implemented in VASPsol.\cite{Mathew2014,Mathew2016} Additional corrections are introduced to correct the solvation errors,\cite{Sundararaman2017} and the reported size-dependence of the free energy.\cite{Bossche2019}  A brief introduction to the computational methods, complete collection of the kinetic parameters, as well as the details of results are available in SI.

\begin{equation}
\ce{H3O+} +\mathrm{e}^- \leftrightarrow \ce{H}_\mathrm{t}^* + \ce{H2O} ~\mbox{(acidic Volmer)}
\end{equation}
\begin{equation}
\ce{H3O+} +\ce{H}_\mathrm{t}^*+\mathrm{e}^- \leftrightarrow \ce{H2} + \ce{H2O} ~\mbox{(acidic Heyrovsky)}
\end{equation}
\begin{equation}
\ce{H2O} +\mathrm{e}^- \leftrightarrow \ce{H}_\mathrm{t}^* + \ce{OH-} ~\mbox{(alkaline Volmer)}
\end{equation}
\begin{equation}
\ce{H2O} +\ce{H}_\mathrm{t}^*+\mathrm{e}^- \leftrightarrow \ce{H2} + \ce{OH-} ~\mbox{(alkaline Heyrovsky)}
\end{equation}
\begin{equation}
2\ce{H}_\mathrm{t}^* \leftrightarrow \ce{H2}~\mbox{(Tafel)}
\end{equation}
\begin{equation}
\ce{H}_\mathrm{t}^* \leftrightarrow \ce{H}_\mathrm{f}^*~\mbox{(Transformation)}
\end{equation}

\begin{figure}
	\includegraphics[width=0.48\textwidth] {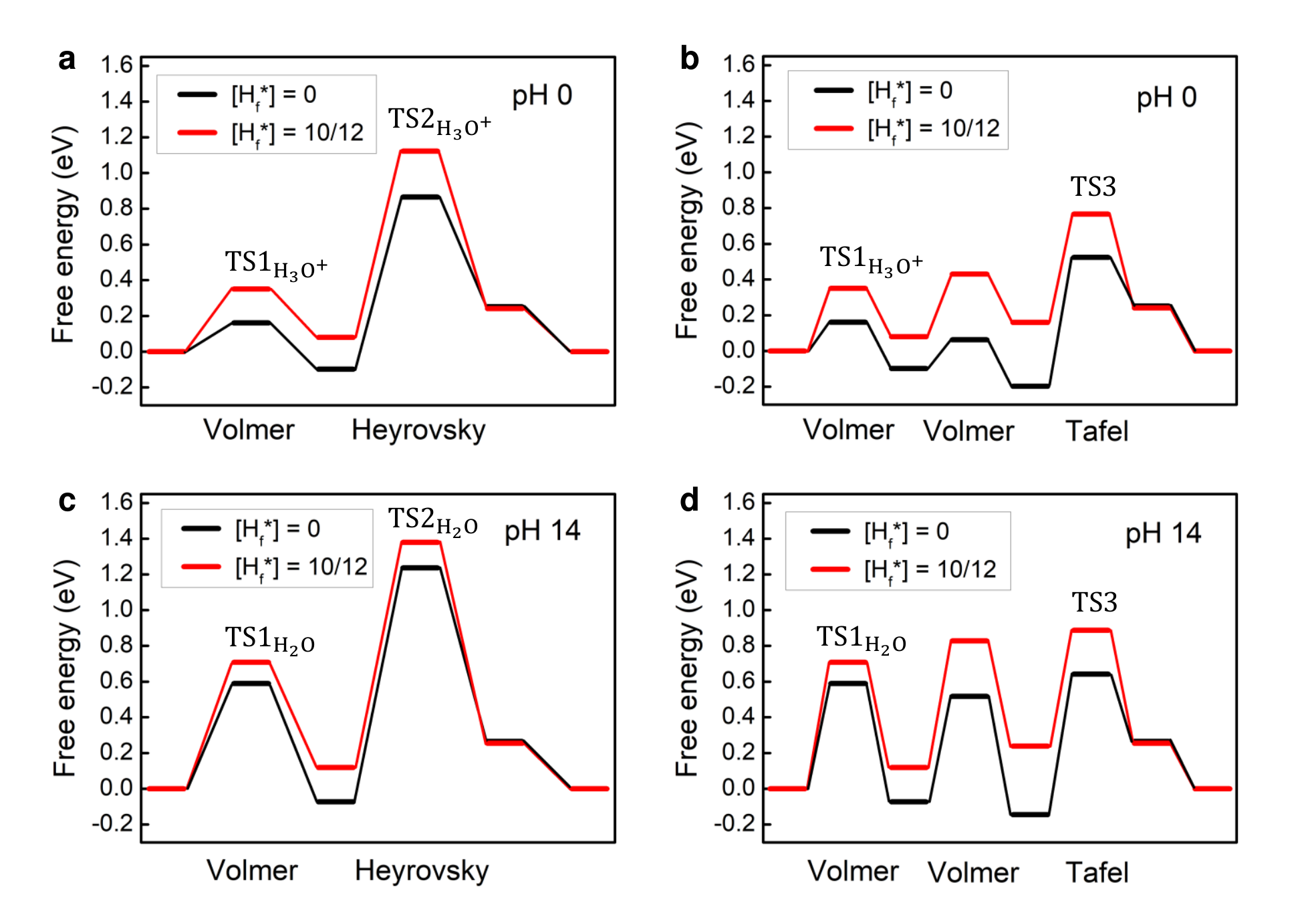}
	\caption{The free energy diagram for Volmer-Heyrovsky and Volmer-Tafel routes at the thermodynamic reversible potentials. Results for clean and $10/12$ monolayer (ML) \ce{H}$_\mathrm{f}^*$  pre-occupied surfaces are displayed.}
	\label{fig:GU}
\end{figure}

Listed above are the six elementary steps in the full pH range modeling of the HER/HOR processes. Besides the regularly investigated Volmer, Heyrovsky, as well as Tafel steps, transformation between two hydrogen states \ce{H}$_\mathrm{t}^*$ (on-top) and \ce{H}$_\mathrm{f}^*$ (fcc) is also included. Figure \ref{fig:GU} shows the four calculated free energy diagrams for the Volmer-Heyrovsky and Volmer-Tafel reaction pathways in both acidic (pH 0) and alkaline (pH 14) conditions. The forward and reverse activation barriers of the Heyrovsky step are found to be always much higher than that of the competing Tafel step, whenever in acidic or alkaline environments, indicating the prevailing of the Volmer-Tafel route over the Volmer-Heyrovsky route in the full pH range. The alkaline process is different from the acidic pathway in the  Volmer step, since the \ce{O-H} bond is stronger in \ce{H2O} than in \ce{H3O+}, which lead to a much higher activation barrier (e.g. 0.428 eV on clean surface) in alkaline Volmer step. Besides, hydrogen coverage on Pt(111) may also considerably affect the kinetics, for an over-crowded surface weakens the hydrogen binding, which should accelerate the forward Tafel and reverse Volmer steps and vice versa, in agreement with the Br\o nsted-Evans-Polanyi (BEP) relation.\cite{Bligaard2004} 

\begin{figure}
	\includegraphics[width=0.48\textwidth] {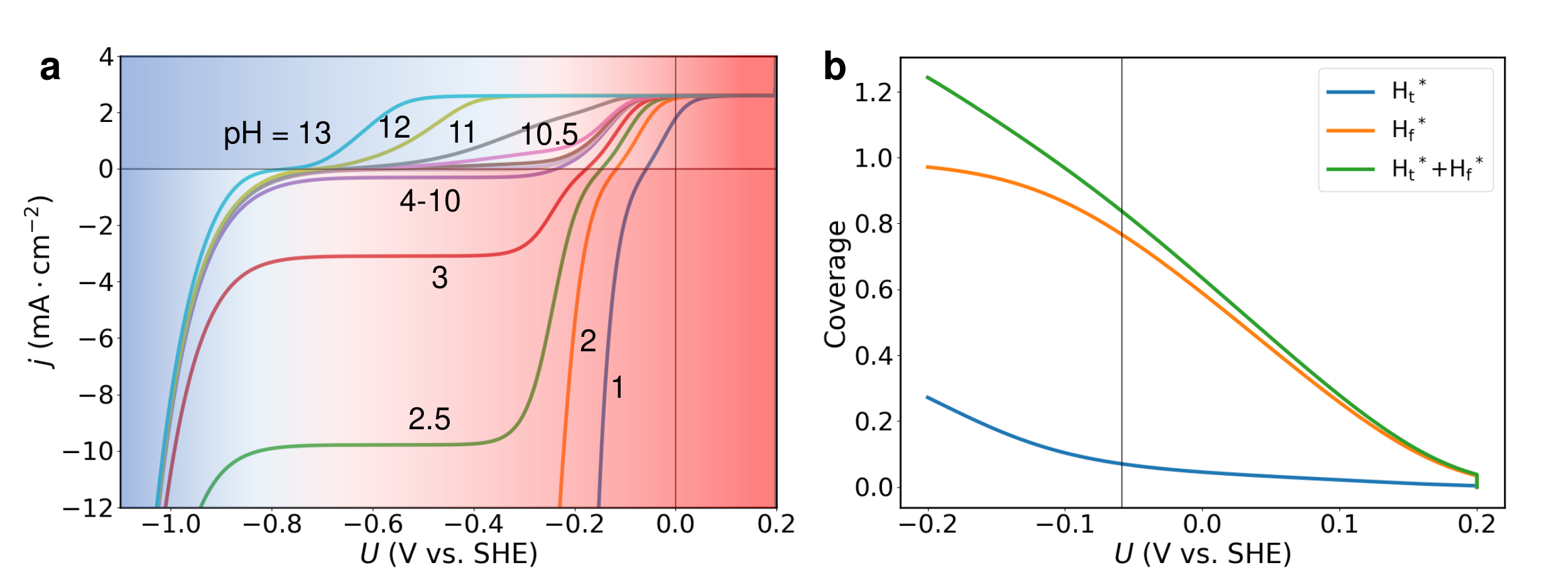}
	\caption{(a) Simulated full pH range polarization curves in \ce{H2}-saturated electrolyte at a rotation rate of 1600 r.p.m. and a sweep rate of 50 mV/s. (b) Plot of the hydrogen coverage vs. potential in acidic condition (pH 1).}
	\label{fig:jE}
\end{figure}

The simulated current-potential polarization curves displayed in Figure \ref{fig:jE}a agree well with the experimental results from Markovic's group,\cite{Strmcnik2013} after a few minor adjustments to the kinetic parameters. Considering the inaccuracy of present generation of DFT methods, such kind of adjustment is usually necessary to reach acceptable coincidence between numerical simulations and experimental observations. As marked in Table S3,  the activation barrier for Tafel step is adjusted by -0.04 eV at 10/12 ML,  and that for alkaline Volmer step is by 0.07 eV at 0 ML and 0.03 eV at 10/12 ML. The self-ionization of \ce{H2O} is neglected, due to that using documented rate constants of ionization/neutralization will result in severe deviation of current curves in the pH range from 2.5 to 4 (Figure S3), which may be attributed to the unusual dynamics of \ce{H+}/\ce{OH-} within the electrical double layer.\cite{Stuve2012,Grozovski2017}

As shown in Figure \ref{fig:jE}a, the calculated polarization curves are divided into the acidic and alkaline current branches. In the HER region, the acidic and alkaline mechanism regions are isolated by the plateau in each curve, characterizing the \ce{H+} diffusion-control kinetics. Above pH 4, the plateau current declines quickly to almost zero, which indicates the reactant of HER is switched from \ce{H3O+} to \ce{H2O} in approaching neutral pH. 
In the HOR region, the pH-independent plateaux around 2 mA/cm$^2$ are attributed to the diffusion control of \ce{H2}, while the expected pH-dependent plateaux of \ce{OH-} diffusion control which should have appeared around pH 10.5 are absent, as a result of the intervention of acidic HOR process. Figure \ref{fig:jE}b presents a linear relationship between the total hydrogen coverage and the applied potential at pH 1, which is ca. 0.63 ML at 0 V vs.\ SHE, compared to the experimental value of 0.66 ML.\cite{Markovic2002} As \ce{H}$_\mathrm{f}^*$  is more stable than \ce{H}$_\mathrm{t}^*$ (within 0.1 eV), it predominates in the hydrogen adsorption in HOR region, thus HER is initiated on the surface highly covered with \ce{H}$_\mathrm{f}^*$ (ca. 0.77 ML). Due to the limited vacancy for \ce{H}$_\mathrm{f}^*$ adsorption, the coverage of \ce{H}$_\mathrm{t}^*$ that formed from the \ce{H3O+} reduction increases superlinearly with the negatively increasing potential, which shows a good agreement with the near-exponential growth of infrared (IR) absorption spectra of \ce{H}$_\mathrm{t}^*$.\cite{Kunimatsu2006} 

The concept of the degree of rate control (DRC) was developed by Campbell to quantify the impact of the free energy perturbation $\delta G_i$ of a specific species $i$ on the total reaction rate, which has been widely adopted to identify the rate-controlling transition states (TS) and intermediates.\cite{Campbell2017} For the convenience in electrochemistry, we slightly modify the original DRC equation to express the degree of control on current density $j$,

\begin{equation}
\mathrm{DRC}_i = -\frac{RT}{j}\left(\frac{\partial j}{\partial G_i}\right)_{G_{k\ne i}} = -RT \left(\frac{\partial \ln \vert j \vert}{\partial G_i}\right)_{G_{k\ne i}}
\end{equation}	

\begin{figure}[H]
	\includegraphics[width=0.48\textwidth] {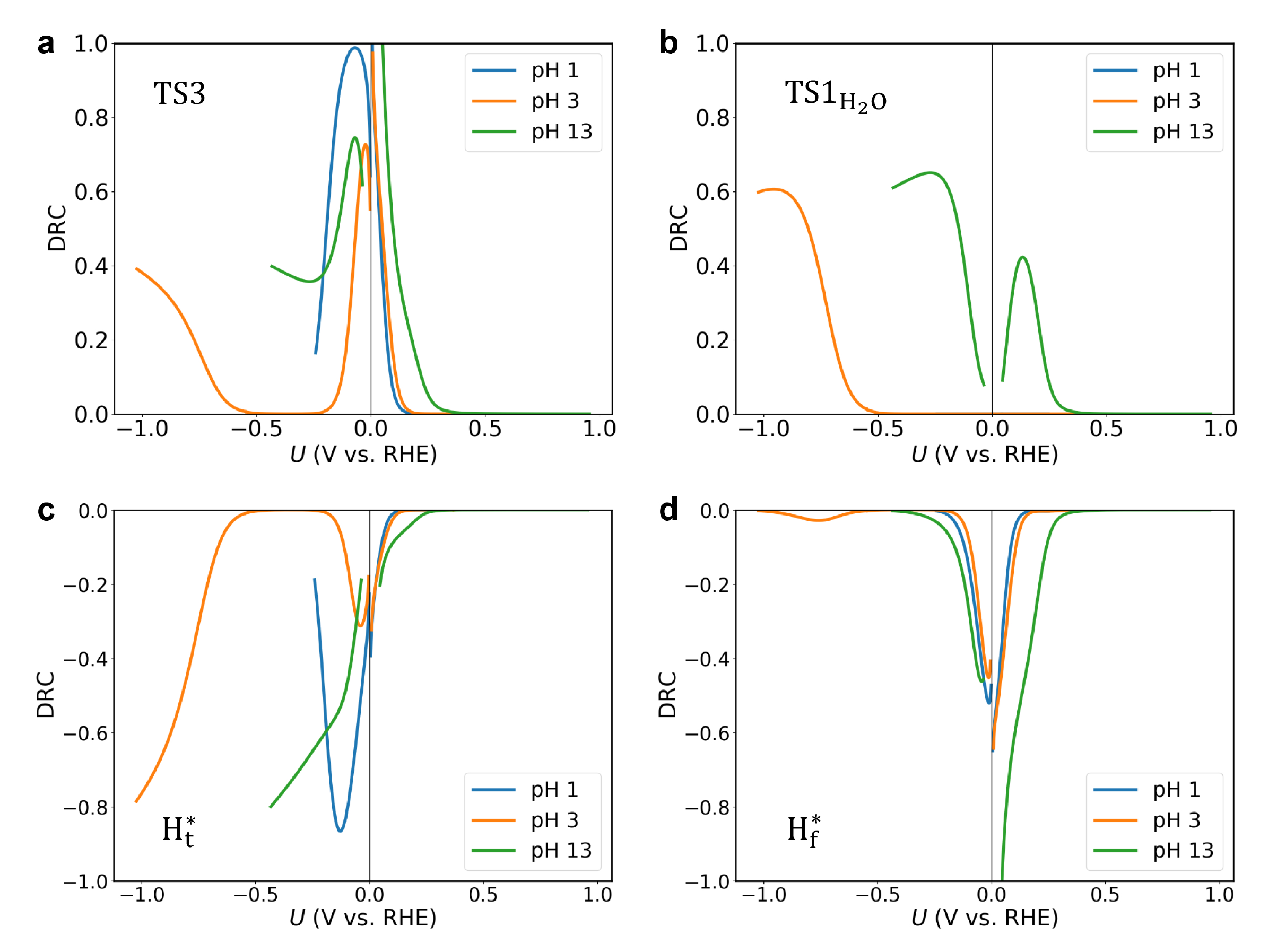}
	\caption{DRC curves for the key species in HER/HOR on Pt (111), including two transition states in (a) Tafel and (b) alkaline Volmer steps, and two intermediates, (c) on-top \ce{H}$_\mathrm{t}^*$ and (d) fcc \ce{H}$_\mathrm{f}^*$. }	\label{fig:DRC}
\end{figure}

\noindent The DRC for transition states are usually positive values (Figure \ref{fig:DRC}a\&b), which means stabilizing a TS will increase the current density. Contrarily, stabilizing an intermediate will often slow down the reaction, corresponding to a negative value (Figure \ref{fig:DRC}c\&d). DRC analysis provides a more comprehensive way to explore the rate-determining step (RDS) than the widely adopted Tafel slope analysis, offering profound information on the evolution of RDS throughout the ranges of potential and pH. Technically, each $\mathrm{DRC}_i$ is computed with the finite-difference method, setting $\delta G_i$ as 0.001 eV. Beside those displayed in Figure \ref{fig:DRC}, DRC plots for the remaining species are collected in SI.

Figure \ref{fig:DRC}a\&b illustrate the effect of pH on the DRC of the transition states in Tafel and alkaline Volmer steps. At pH 1 and within small overpotentials, the calculated DRCs of the Tafel step present to be close to 1.0, indicating an exclusive control on the kinetics of HER/HOR, which is in accordance with earlier experimental and theoretical studies.\cite{Seto1987,LedezmaYanez2017,Skulason2010} At pH 3, the zero value of DRCs indicates the \ce{H+} diffusion control in the HER region between -0.2 and -0.5 V. Out of this potential range, the HER process follows acidic or alkaline mechanisms respectively in lower or higher overpotential regions. In alkaline condition,  earlier experimental studies of HER/HOR in \ce{H2} saturated electrolyte on Pt(111) found that the Tafel slope increased continuously with increasing overpotentials, covering the distinctive slopes of Volmer, Tafel and Heyrovsky steps, so that the RDS can not be unambiguously assigned.\cite{Barber1999,Schmidt2002} However, a more recent experimental study suggested the unique control of Volmer step in the alkaline HER current, according to the Tafel plot obtained under argon atmosphere,\cite{LedezmaYanez2017}  while that in \ce{H2} saturated electrolyte seemed a similar feature with previous experiments. Despite the disagreement in the experimental results, the DRC curves for pH 13 show clearly an incomplete transition of the RDS from Tafel step to Volmer step, indicating the alkaline HER/HOR current is controlled jointly by both steps in the kinetics-controlled region. 

Analysis on the DRC profiles of \ce{H}$_\mathrm{t}^*$ and \ce{H}$_\mathrm{f}^*$ can reveal their respective roles in controlling the total reaction rate, which can enhance our understanding on the hydrogen intermediates acting as an activity descriptor in HER/HOR processes. Obviously, DRC values of \ce{H}$_\mathrm{t}^*$ in HER potential region are much larger than that in HOR region in the whole investigated pH conditions, while those of \ce{H}$_\mathrm{f}^*$ behave in an opposite way. This unusual discovery can be rationalized, considering the rapid transformation between \ce{H}$_\mathrm{t}^*$ and \ce{H}$_\mathrm{f}^*$. Above the thermodynamic reversible potential, the hydrogen atoms reside predominantly on fcc sites, and play the role of a reactive species in the Volmer step, due to the short lifetime of \ce{H}$_\mathrm{t}^*$. On the other hand, when the overpotentials of HER applied, considerable amount of \ce{H}$_\mathrm{t}^*$ has been produced to take over the role of \ce{H}$_\mathrm{f}^*$, while the later becomes a spectator on the surface. Therefore, it is interesting to think about the possibility of tuning two kinds of adsorptions with different strategies in optimizing HER/HOR performance.
 
\begin{figure}
	\includegraphics[width=0.48\textwidth] {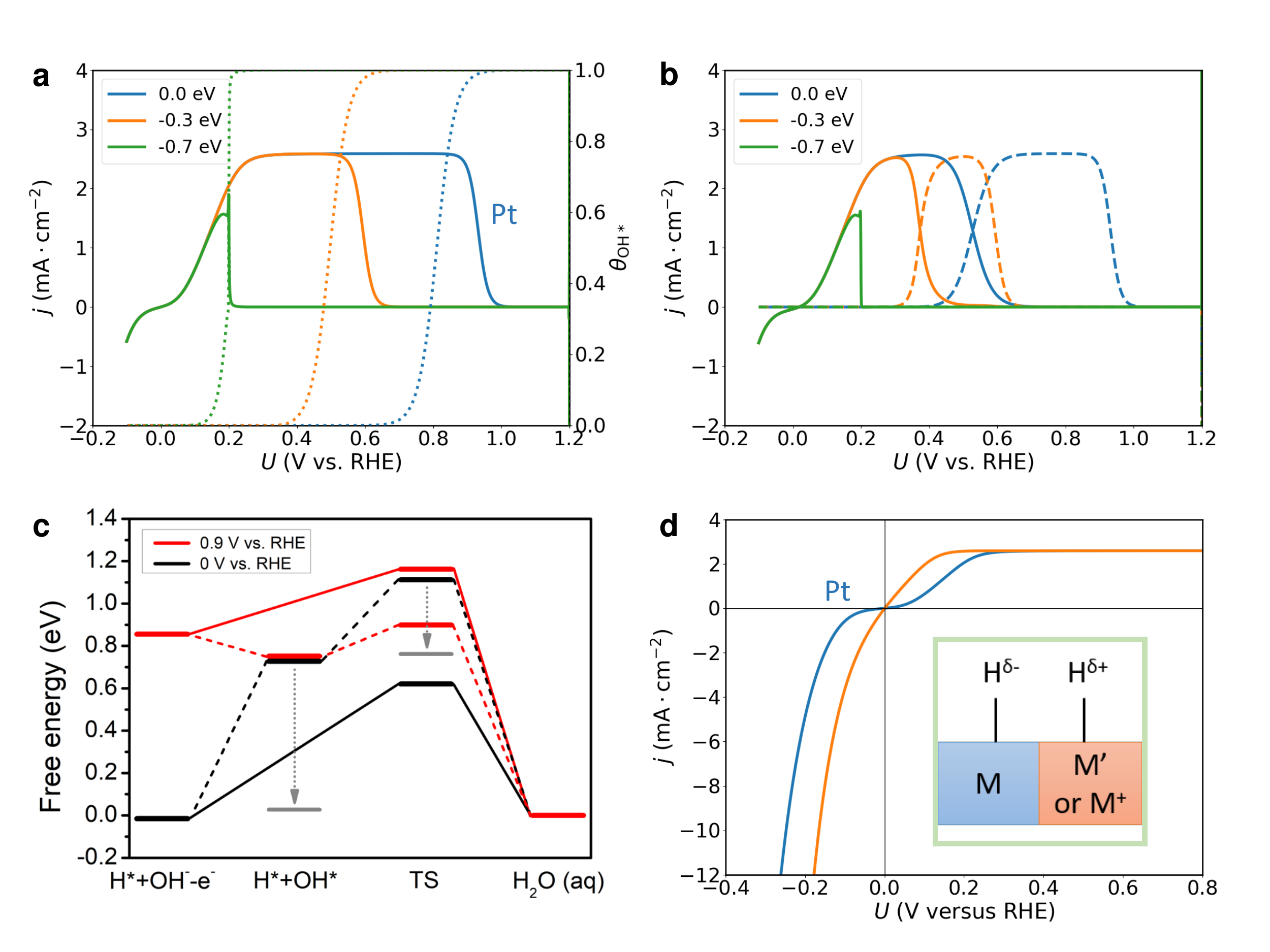}
	\caption{(a) The current-potential polarization curves (pH 13) calculated with varying \ce{OH}$^*$ stabilities, with \ce{OH}$^*$ coverage shown in the dotted curves. (b) The partial current density for \ce{OH}$^*$ (dotted line) and \ce{OH-} (full line) mechanism. (c) Free energy diagram of the competing \ce{OH-} (full lines) and \ce{OH}$^*$ (dashed lines) mechanisms in HOR.  (d) An illustration of tuning the reaction kinetics through stabilizing the transition sate of the Tafel step by 0.1 eV. }
	\label{fig:oh}
\end{figure}

As shown in Figure \ref{fig:oh}a\&b, two model systems with stronger metal-\ce{OH}$^*$ binding strength than Pt (111) surface are computed to look into the role of \ce{OH}$^*$ in alkaline HOR.  The reaction barriers relevant to \ce{OH}$^*$ are adjusted following the BEP principle with a slope of 0.5. For doing this, two extra elementary steps are introduced to account for the oxidative adsorption of \ce{OH-} as well as the the combination of \ce{H}$^*$ and \ce{OH}$^*$.

\begin{equation}
\ce{OH-} \leftrightarrow \ce{OH}^* + \mathrm{e}^-
\end{equation}
\begin{equation}
\ce{H}_\mathrm{t}^* +\ce{OH}^* \leftrightarrow  \ce{H2O} 
\end{equation}

Increasing the binding strength of \ce{OH}$^*$ is found to impose no effect on the onset potential of HOR, instead, it will greatly influence the current-breakdown potential, which is obviously coincident with the desorption potential of \ce{OH}$^*$. Examination of the \ce{OH}$^*$ coverage (Figure \ref{fig:oh}a), as well as the partial currents contributed from \ce{OH}$^*$-mediated and direct \ce{OH-} oxidation (Figure \ref{fig:oh}b) can clearly interpret whether \ce{OH}$^*$ takes a part in affecting the kinetics of alkaline HOR. \ce{OH}$^*$ is found to play the role of a reactive species in a certain range of potential,  starting from the \ce{OH}$^*$ desorption potential, down to somewhere the direct \ce{OH-} oxidation mechanism takes over the \ce{OH}$^*$ mechanism, due to the fast decrease of the \ce{OH}$^*$ coverage and stabilization of TS in the \ce{OH-} oxidation step with decreasing potential. Approaching the HOR onset region, the \ce{OH}$^*$ coverage becomes negligible, together with its influence on HOR kinetics.
Figure \ref{fig:oh}c presents a schematic diagram of the relative free energies of the reactive species on Pt(111) computed at 0 and 0.9 V vs. RHE. At 0 V, \ce{OH-} is found to be much more stable than \ce{OH}$^*$, such that the oxidation path through \ce{OH}$^*$ is shut down. Tuning of \ce{OH}$^*$ towards a more stable binding (the grey line) will not alter the oxidation mechanism, if only the transition state of which remains above that of \ce{OH-}. While at 0.9 V, \ce{OH-} is considerably destabilized while \ce{OH}$^*$ mainly keeps its stability, such that HOR can possibly proceed through the \ce{OH}$^*$-mediated path. 

The above theoretical analyses depend on a simplified model assuming \ce{H}$^*$ binding strength is constant with varying \ce{OH}$^*$ stabilities. In real systems, the stabilities of \ce{H}$^*$ and \ce{OH}$^*$ can be internally correlated. Therefore, the previously reported impact of \ce{OH}$^*$ binding strength on HOR activity might also be originated from the concurrently varying stability of \ce{H}$^*$. In some other systems, taking \ce{H}$^*$ as the sole descriptor has been proved to be effective.\cite{Sheng2015,Wang2015,Lu2017} At least for alkaline HOR processes on Pt(111) and systems that share the same mechanism, strengthening the \ce{OH}$^*$ binding is not only unnecessary for accelerating the alkaline HOR, but also could cause a lower break-down potential and suppress the current density (Figure \ref{fig:oh}a), just like the case on Ru(0001) surface.\cite{Strmcnik2013}

Based on the theoretical simulation and DRC analyses, the dramatic difference of kinetic performance for alkaline HER/HOR is due to that the activation barrier of the Volmer step is considerably increased while that of the Tafel step remains essentially unchanged, as compared with the acidic process. The mechanism switching in Volmer step can be taken as the breaking of BEP principle, since the two conditions share the same intermediate state \ce{H}$^*$, but the activation barrier is obviously different. Similarly, instead of relying on the modulation of HBE, it is helpful to think about optimizing the alkaline HER/HOR activity through stabilizing the TS of the Tafel step and maintaining the stability of \ce{H}$^*$, which could considerably reduce the overpotential, as shown in Figure \ref{fig:oh}d. One possible strategy for stabilizing the TS is to polarize the surface hydrogen couples ($\ce{H}^{*\delta+} +\ce{H}^{*\delta-} \leftrightarrow \ce{H2}^*$), and a similar approach has been proposed in \ce{CO} dimerization.\cite{Xiao2017}

\begin{acknowledgement}
	This work is supported by China NSF (Grant No. 21173116, 21473088) and the Special Program for Applied Research on Super Computation of the NSFC-Guangdong Joint Fund (the second phase) under Grant No. U1501501.
\end{acknowledgement}

\begin{suppinfo}
	Detailed computational schemes for the electrode/electrolyte model, potential-related kinetic parameters, configurations of optimized structure, DRC curves for the non-rate-controlled species.
\end{suppinfo}

\bibliography{ref}

\newpage

\begin{tocentry}
	\includegraphics[height=4.45cm] {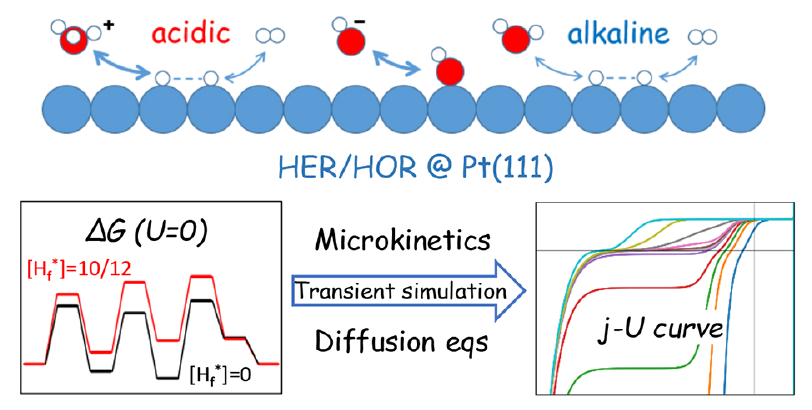}
	\label{fig:For Table of Contents Only}
\end{tocentry}

\end{document}